\documentstyle[12pt]{article}
\begin{document}
\setlength{\baselineskip}{20pt}
\begin{center}
{\large{\bf Hamiltonian structure and Darboux' theorem \\ 
for families of Generalized Lotka-Volterra systems}} \\ \mbox{} \\
{\bf Benito Hern\'{a}ndez--Bermejo \hspace{.5in}
V\'{\i}ctor Fair\'{e}n$^*$}
\end{center}
\noindent {\em Departamento de F\'{\i}sica Fundamental, Universidad Nacional 
de Educaci\'{o}n a Distancia. Senda del Rey S/N, 28040 Madrid, Spain.}

\mbox{}

\begin{center} 
{\bf Abstract}
\end{center}
This work is devoted to the establishment of a Poisson structure for a 
format of equations known as Generalized Lotka-Volterra systems. 
These equations, which include the classical Lotka-Volterra systems as a 
particular case, have been deeply studied in the literature. 
They have been shown to constitute a whole hierarchy of systems, the 
characterization of which is made in the context of simple algebra. 
Our main result is to show that this algebraic structure is completely 
translatable into the Poisson domain. Important Poisson structures features, 
such as the symplectic foliation and the Darboux' canonical representation, 
rise as result of rather simple matrix manipulations. 

\mbox{}

\mbox{}



\noindent {\bf Short Title:} Hamiltonian structure for Generalized LV systems.

\mbox{}

\mbox{}

\noindent $^*$ Corresponding author. Email: vfairen@uned.es

\begin{flushleft}
{\bf I. INTRODUCTION}
\end{flushleft}

Poisson structures$^{1,2 \:}$ (sometimes named {\em generalized Hamiltonian 
structures\/} in the literature) are ubiquitous in all fields of 
Mathematical Physics, from finite-dimensional dynamical systems$^{3-7 \:}$ to 
field theories:$^{8,9 \:}$ Fluid dy\-namics,$^{10,11 \:}$ 
magnetohydrodynamics,$^{11,12 \:}$ plasmas,$^{13-15 \:}$ continuous 
media,$^{15 \:}$ 
condensed matter,$^{16 \:}$ etc. Reformulating a given problem in 
terms of a Poisson structure provides fruitful insight into the behaviour of 
the system, which may take the form of perturbative solutions,$^{17 \:}$ 
nonlinear stability analysis through the energy-Casimir algorithm$^{7,18 \:}$ or 
the energy-momentum method,$^{19 \:}$ bifurcation properties and 
characterization of chaotic dynamics,$^{20 \:}$ integrability 
results,$^{21 \:}$ 
application of reduction of order procedures$^{2,22 \:}$ or explicit 
determination of new solutions.$^{14,23 \:}$ 

In the present work, we shall restrict ourselves to finite-dimensional 
Poisson structures. In terms of local coordinates, a Poisson system defined 
on an $n$-dimensional manifold takes the following form:
\begin{equation}
\label{nham}
    \dot{x}_i = \sum_{j=1}^n J_{ij} \partial _j H \; , \;\:\; i = 1, \ldots , n
\end{equation} 
The smooth, real-valued function $H(x)$ in (\ref{nham}) is a constant of 
motion of the system, which plays the role of Hamiltonian, and the $J_{ij}(x)$ 
are also smooth and real-valued, being the entries of a $n \times n$ 
skew-symmetric structure matrix ${\cal J}$ which verifies the Jacobi 
equations:
\begin{equation}
     \label{jac}
     \sum_{l=1}^n ( J_{li} \partial_l J_{jk} + J_{lj} \partial_l J_{ki} + 
     J_{lk} \partial_l J_{ij} ) = 0 
\end{equation}
Here $ \partial_l $ means $ \partial / \partial x_l$ and indices $i,j,k$ run 
from 1 to $n$. The flow (\ref{nham}) can then be expressed as $\dot{x}_i$ $=$ 
$[x_i,H]$, in terms of the Poisson bracket defined by
\[
   [F,G] = \sum_{i,j=1}^n \frac{ \partial F}{ \partial x_i} J_{ij} 
   \frac{ \partial G}{ \partial x_j} \;\; , 
\]
where $F$ and $G$ are smooth real-valued functions defined on the Poisson 
manifold. Consequently, Poisson structures generalize classical Hamiltonian 
systems, for which ${\cal J}$ is the well-known symplectic matrix. In 
particular, the classical 
restriction to even-dimensional manifolds is not present in Poisson 
systems. However, Poisson dynamics preserves the Hamiltonian character of the 
motion. This is proven by Darboux' theorem,$^{2 \:}$ which states that 
there exist local coordinates in the neighbourhood of every point of the 
Poisson manifold, such that the equations of motion take essentially the 
classical Hamiltonian form. The practical construction of Darboux' 
coordinates is, however, a complicated task in general, which has been 
carried out only for a limited sample of systems.$^{2,24,25 \:}$ 
 
An important question is that of characterizing a given vector field 
not in form (\ref{nham}), as an actual 
Poisson system. In the finite-dimensional case the problem amounts to giving 
a procedure for decomposing (whenever possible) a smooth function $f(x): 
\Omega \subset \mbox{R}^n \longrightarrow \mbox{R}^n$, where $\Omega$ is open, 
as $f(x)= {\cal J}(x) \cdot \nabla H(x)$, where ${\cal J}$ is a solution of 
the nonlinear PDE (\ref{jac}) and $H(x)$ is a real-valued function. This is a 
nontrivial problem to which important efforts have been devoted in past years 
in a variety of approaches.$^{3-8, 25-28 \:}$ The question is well understood 
in the simplest cases ---two and three dimensions--- and the existence and 
determination of at least one Poisson structure is ensured if a first 
integral is known for the system.$^{25,28 \:}$ In higher dimensions the 
situation is by no means so clear, and comparable results are still lacking. 

The main exception to this absence of results in $n$-dimensional systems is, 
to our knowledge, given by the Lotka-Volterra equations (LV from now on). 
They were introduced by Lotka$^{29 \:}$ and Volterra$^{30 \:}$ in chemical and 
biological contexts, respectively, and Volterra himself was already aware of  
the (classical) Hamiltonian nature of some LV systems. However, the main and 
more systematic advances are due to Kerner,$^{31 \:}$ who 
developed a biological analog of classical statistical mechanics for 
predator-prey systems. More recently, the attention has turned to the search 
of Poisson structures for more general LV models, since Poisson structures 
generalize classical Hamiltonian systems while retaining the Hamiltonian 
nature of the dynamics ---in this sense, the Poisson structure of many two 
and three-dimensional LV systems has been established,$^{3,25,27,28 \:}$ and 
all three-dimensional biHamiltonian LV systems have been classified by 
Plank.$^{5 \:}$ Also, in the domain of $n$-dimensional LV flows, a first  
tentative classification of Poisson structures has been carried out.$^{6 \:}$ 

Both the relevance of LV equations and the importance of finding their 
Hamiltonian or Poisson representations, trascends the fact that LV models are 
appropriate in describing many problems in Biology, Chemistry or Physics. 
Cair\'{o} and Feix,$^{32 \:}$ for example, refer to a fairly long list of 
systems modelled by LV equations ---their ubiquity has even prompted Peschel 
and Mende$^{33 \:}$ to head their book on the issue with the title: {\em Do 
we live in a Volterra World?\/} 

Actually, the significance of LV equations goes beyond a strict modelling 
context, because they have been shown to be canonical representatives of 
infinite families, or classes, of very general flows,$^{33,34 \:}$ 
to which, following Brenig, we shall refer as Generalized Lotka-Volterra 
(GLV) systems. There is a whole formalism associated to the GLV 
equations.$^{33-36 \:}$ As we shall see later, the most relevant feature of 
this formalism is that of permitting the analysis and interpretation of 
certain properties of the vector field in purely algebraic terms. 

Our purpose in this article is to demonstrate how these algebraic properties 
are of fundamental importance in understanding the Poisson structure of 
LV and GLV models. We shall investigate Poisson structures for GLV families 
of systems. They include as particular elements all LV models which have been 
the object of interest in the literature in relation to Poisson structures 
(in works of Volterra, Kerner and Plank). We shall show that the algebraic 
GLV matrix properties can be translated into the Poisson context and acquire 
a new significance. The reverse is also true, thus defining a close 
connection between GLV algebraic properties and the Poisson structure of the 
system. 

Implementing the structure of the GLV formalism on its Poisson counterpart 
has very interesting consequences. First of all, we are able to include in 
our scheme systems which are more general than the LV ones. Second, we can 
take into account larger sets of LV flows than those studied by previous 
approaches. For example, we are neither limited to LV systems of even 
dimension, nor cases with a unique fixed point. These are two common 
restrictions often imposed in the literature,$^{6,37 \:}$ which we obviate at 
once. Third, we are able to capture important phase-space features in terms 
of simple properties of constant matrices. Finally, our approach leads to an 
algorithmic reduction to the Darboux' form for the equations. And last but 
not least, our construction is always global. 

\mbox{}

\begin{flushleft}
{\bf II. OVERVIEW OF THE GLV FORMALISM}
\end{flushleft}

We proceed now to briefly summarize the main features of the GLV formalism. 
We refer to the reader interested in a more detailed exposition to the 
original references.$^{33-36 \:}$ 

\pagebreak
{\em Definition 2.1:\/} A GLV system is a set of ordinary differential equations 
which is defined in the real positive orthant and complies to the form: 
\begin{equation}
   \dot{x}_i = x_i(\lambda _{i} + \sum_{j=1}^{m}A_{ij}\prod_{k=
      1}^{n}x_k^{B_{jk}}) , \;\:\;\: i = 1 \ldots n 
   \label{eq:glv}
\end{equation}
where $n$ and $m$ are positive integers, $m \geq n$, and $A$, $B$ and 
$\lambda$ are $n \times m$, $m \times n$ and $n \times 1$ real matrices, 
respectively. 

The $m$ nonlinear terms of the right-hand side of (\ref{eq:glv}) are 
usually known as {\em quasimonomials.\/}  
Sometimes we shall group all the coefficients of $A$ and $\lambda$ in a 
single, composite matrix, $M = ( \lambda \mid A)$. We shall assume that 
matrix $B$ is of maximal rank. This is a standard case to which every GLV 
system can be reduced.$^{36 \:}$ Notice also that the well-known LV equations 
\[
   \dot{x}_i = x_i(\lambda _{i} + \sum_{j=1}^{n}A_{ij}x_j) , \;\:\;\: 
   i = 1 \ldots n 
\]
are a particular case of (\ref{eq:glv}) where $m=n$ and $B$ is the $n \times 
n$ identity matrix.

System (\ref{eq:glv}) is form-invariant under quasimonomial transformations 
(or QMTs from now on):
\begin{equation}
   x_i = \prod_{k=1}^{n} y_k^{C_{ik}} , \;\: i=1,\ldots ,n \; , 
   \;\:\; \mid C \mid \neq 0
\label{bec}
\end{equation}
Under (\ref{bec}), matrices $B, A, 
\lambda$ and $M$ change to $B' = B \cdot C$, $A' = C^{-1} \cdot A$, 
$\lambda ' = C^{-1} \cdot \lambda $ and $M' = C^{-1} \cdot M$, respectively, 
but the GLV format is preserved. Obviously, all GLV systems which can be 
connected through QMTs share the value of the product $B \cdot M$. These 
families of systems are, in fact, classes of equivalence, the product 
$B \cdot M$ being a class invariant. The QMTs are diffeomorphisms defined 
in the positive orthant, and are orientation-preserving iff 
$ \mid C \mid > 0$. Consequently, QMTs preserve the topology of the phase 
portrait modulo an inversion. 

{\em Definition 2.2:\/} A GLV class of equivalence for which rank($M$) $=r$, 
and whose members are $n$-dimensional and have $m$ quasimonomials, is denoted 
as an ($r,n,m$)-class. 

The kind of manipulations in which we shall be interested later will 
transform a GLV system into another one belonging to the same or, eventually,  
to a different class. However, these manipulations will affect neither $r$ 
nor $m$, but may change $n$. We shall always operate, however, in the range 
$r \leq n \leq m$. Obviously, a QMT does not modify anyone of these three 
indexes. 

If $m=n$, we can perform a QMT of matrix $C = B^{-1}$. The result is another 
flow for which $B' = I_n$, that is, an LV system. Such a system can be 
taken as the canonical representative of the GLV class of equivalence. 

In the complementary case $m > n$, there is no LV system inside the class of 
equivalence. However, the reduction to the LV form is possible if we perform 
an embedding, just by adding new variables to system (\ref{eq:glv}). 

{\em Definition 2.3:\/} We define a $p$-embedding as the result of adding to 
a GLV system $p$ new variables in the following way:
\[
    \dot{x}_i = 0 \; , \;\; x_i(0)= \alpha _i >0 \; , \;\; i= n+1, \ldots , 
    n+p \; , \;\; 1 \leq p \leq m-n
\]

Let $A$, $B$, and $\lambda$ be the matrices of the original GLV system. 
The $p$-embedded system is also GLV, and its characteristic matrices are:
\begin{equation}
\label{bemb}
     \tilde{B} = ( B \mid B^*_{m \times p}) \; , \;\;\: 
     \tilde{ \lambda } = \left( \begin{array}{c} 
              \lambda \\ O_{p \times 1} 
                         \end{array} \right) \; , \;\;\: 
     \tilde{ A } = \left( \begin{array}{c} 
              A \cdot E \\ O_{p \times m}
                   \end{array} \right) \; ,
\end{equation}
where 
\begin{equation}
\label{e1}
E= \mbox{diag}(e_1, \ldots , e_m) \; , \;\;\; 
     e_j = \left( \prod_{k=n+1}^{n+p} \alpha _k^{\tilde{B}_{jk}} 
     \right)^{-1} \; ,\; \; \; j=1, \ldots ,m  
\end{equation}
In (\ref{bemb}), $O$ denotes a submatrix of null entries, while 
$B^*_{m \times p}$ has arbitrary real entries appropriately chosen for 
$\tilde{B}$ to be of maximal rank. The subscripts such as $_{m \times p}$ 
indicate the size of the corresponding submatrix ($B^*$ in this case); we 
shall maintain this notation henceforth.

Notice how the previous operation transforms a GLV system from an 
($r,n,$ $m$)-class, with $m>n$, into a GLV system belonging to an 
($r,n+p,m$)-class, with $1 \leq p \leq m-n$. The embedded system is 
topologically equivalent to the original one in the manifold 
$x_i= \alpha _i, i= n+1, \ldots , n+p$. 

For fixed $p$, infinitely many different ($r,n+p,m$)-classes can be reached 
by means of an embedding, depending on the entries of matrix $E$. Also, in 
the particular case $p=m-n$ the target system belongs to an ($r,m,m$)-class, 
in which an LV representative can be reached by a QMT of 
matrix $C$ $=$ $\tilde{B}^{-1}$ (since rank($\tilde{B}$) $=m$). Thus, the 
recasting of the original GLV flow into LV form is always possible. For 
this $m$-dimensional LV system, however, rank($\tilde{M}_{LV}$) $<m$, and is 
not maximum. Whenever this happens in a GLV system, it indicates the 
existence of quasimonomial constants of motion, as the following proposition 
shows:

{\em Proposition 2.4:\/} In a GLV system (\ref{eq:glv}), rank($M$) $= r < n$ 
if and only if there exist $(n-r)$ functionally independent constants of the 
motion which are time-independent and have quasimonomial form. 

{\em Proof:\/} We can assume, without loss of generality, that the $r$ first 
rows of $M$ are the linearly independent ones. Then, there exist real 
constants $\gamma _{ki}$, with $i= 1, \ldots ,r$, and  $k = r+1, \ldots ,n$, 
such that: 
\[
   M_{kl} = \sum_{i=1}^{r} \gamma_{ki} M_{il} \;, \;\: 
   \forall \; l = 1 , \ldots , m+1 
\]
From (\ref{eq:glv}), we arrive at:
\[   
   \frac{\dot x_k}{x_k} = \sum_{i=1}^{r} \gamma_{ki} \frac{\dot x_i}{x_i} \;.
\]
After a simple integration this leads to the set of $(n-r)$ constants of 
motion:
\[
   x_k^{-1} \prod_{i=1}^{r} x_i^{\gamma _{ki}} = c_k \; , 
\]
where the $c_k $ are real constants given by the initial conditions. 
The functional independence holds immediately from simple evaluation of the 
Jacobian. The proof in the opposite sense is straightforward after this. 
Q.E.D.

Therefore, a $p$-embedding in the $m>n$ case introduces $p$ quasimonomial 
constants of motion, which are obviously form-invariant under QMTs. This 
invariance implies that the quasimonomial constants of motion can always be 
decoupled from a GLV system by means of an appropriate QMT. When this is 
done, what we are doing is to reverse the $p$-embedding procedure, actually. 
The first step to show this is the following result: 

{\em Proposition 2.5:\/} Let $A^*$, $B^*$ and $\lambda ^*$ be the matrices 
of a GLV system belonging to an ($r,n+p,m$)-class, where $r \leq n$ and 
$0<p \leq m-n$. Then there exists a quasimonomial transformation that leads 
to an $(n+p)$-dimensional GLV system of matrices 
\begin{equation}
\label{mcc}
         \lambda = \left( \begin{array}{c}
                          \bar{\lambda} \\ O_{p \times 1}
                   \end{array} \right) \; , \;\;
         A = \left( \begin{array}{c}
               \bar{A} \\ O_{p \times m}
             \end{array} \right) 
\end{equation} 

{\em Proof:\/} We shall omit it, since it is based on simple matrix algebra  
properties. 

In (\ref{mcc}), we have decoupled the final $p$ components of the vector 
field: Let $x_1, \ldots , x_{n+p}$ be the variables of the system of matrices 
(\ref{mcc}), and let $x_{n+i}(0)= \alpha_{n+i}>0$, $i= 1, \ldots ,p$, be the 
initial conditions of the decoupled variables. Let us also write 
$B=(\bar{B} \mid \bar{B}'_{m \times p})$ for the matrix of exponents of this 
system. Then, when we restrict the dynamics to an $n$-dimensional flow, the 
result is another GLV system from an ($r,n,m$)-class, which is characterized 
by three matrices $\hat{B}$, $\hat{\lambda}$ and $\hat{A}$, given by:
\[
    \hat{B}= \bar{B} \; , \; \; \hat{\lambda} = \bar{\lambda} \; , \; \; 
    \hat{A}= \bar{A} \cdot E^{-1} \; , 
\]
where again 
\begin{equation}
\label{e2}
E= \mbox{diag}(e_1, \ldots ,e_m) \; , \;\;\; 
     e_j = \left( \prod_{k=n+1}^{n+p} \alpha _k^{B_{jk}} \right)^{-1}
     \; ,\; \; \; j=1, \ldots ,m
\end{equation}

{\em Definition 2.6:\/} The previous operation transforming a GLV system in 
an ($r,n+p,m$)-class, with $r \leq n$ and $0<p \leq m-n$, into a GLV system 
belonging to an ($r,n,m$)-class, is called a $p$-decoupling.

Even for fixed $n$, there are again infinite possible target classes, due to 
the arbitrariness in the initial conditions represented by $E$. In any case, 
there is obviously a conservation of the topological properties of the flow 
in the process. It is also clear that, in the especial case in which we 
choose $n=r$, we have the maximum reduction possible by means of this method; 
otherwise the decoupling is partial. In either case, the simplification is 
possible because we are, in fact, restricting the dynamics to the level 
surfaces of quasimonomial constants of motion. 

To summarize, we have a multilevel structure of ($r,n,m$)-classes of 
equivalence, with $n$ ranging in the interval $r \leq n \leq m$. We can 
transform freely every GLV system inside this scheme by means of the QMTs and 
the two basic ---and opposite--- operations: $p$-embeddings and 
$p$-decouplings, which proceed by the introduction of quasimonomial first 
integrals, or by the restriction of the system dynamics to their level 
surfaces, respectively. 

\mbox{}

\pagebreak
\begin{flushleft}
{\bf III. GLV FAMILIES OF POISSON SYSTEMS}
\end{flushleft}

We start by characterizing the systems of interest. In what follows, the 
superscript $^T$ denotes the transpose of a matrix. 

{\bf Theorem 3.1:} Let us consider a GLV system of the form (\ref{eq:glv}) 
such that 
\begin{equation}
\label{kdl}
   \lambda = K \cdot L \; , \;\:\;\; A = K \cdot B^T \cdot D \; ,
\end{equation}
with $K$, $L$ and $D$ matrices of real entries, where $K$ is $n \times n$ and 
skew-symmetric; $L$ is $n \times 1$; and $D$ is $m \times m$, diagonal and of 
maximal rank. Then the system has a constant of motion of the form:
\begin{equation}
\label{H}
   H= \sum_{i=1}^m D_{ii} \prod_{k=1}^n x_k^{B_{ik}} + \sum_{j=1}^n L_j 
   \ln (x_j) 
\end{equation}
Moreover, the system is Poisson with Hamiltonian $H$.

{\em Proof:\/} The GLV flow complies to the format $\dot{x}={\cal J} \cdot 
\nabla H$, where the Hamiltonian is smooth in the positive orthant and given 
by $H$ in (\ref{H}), while ${\cal J}$ is the smooth matrix 
\begin{equation}
\label{J}
   {\cal J} = X \cdot K \cdot X \; , \;\;\; X= \mbox{diag}(x_1, \ldots ,x_n)
\end{equation}
That $H$ is the Hamiltonian implies that it is a constant of motion. Q.E.D. 

Notice that the first part of the Hamiltonian is associated to the $m$ 
quasimonomials of the GLV vector field (in fact, it is a linear combination 
of them), while the logarithmic terms are closely connected to the linear 
contributions. The observation that a matrix of the form $X \cdot K \cdot X$ 
is a structure matrix iff $K^T=-K$ is due to Plank.$^{6 \:}$ From now on, 
we shall denote the systems described by Theorem 3.1 as GLV-Poisson (GLVP).

{\em Proposition 3.2:\/} The Poisson structure of GLVP systems is 
form-invariant under a QMT. After a QMT of matrix $C$, the characteristic 
matrices of the transformed Poisson structure are:
\[
   K' = C^{-1} \cdot K \cdot (C^{-1})^T \; , \;\;\: 
   L' = C^T \cdot L \; , \;\;\: 
   D' = D
\]
In particular, both the Hamiltonian and the structure matrix are 
form-invariant under QMTs.

{\em Proof:\/} The simplest proof is the algebraic one. After a QMT we have: 
\begin{eqnarray*}
& \lambda ' = C^{-1} \cdot K \cdot L = C^{-1} \cdot K \cdot (C^{-1})^T \cdot 
C^T \cdot L = K ' \cdot L' & \\
& A'=C^{-1} \cdot K \cdot B^T \cdot D = K' \cdot (B')^T \cdot D' & 
\end{eqnarray*}
and $D' =D$. Then, from Theorem 3.1 the new system is also GLVP, and its 
structure matrix and Hamiltonian are, respectively, ${\cal J'}=Y \cdot K' 
\cdot Y$, and 
\[
   H'= \sum_{i=1}^m D'_{ii} \prod_{k=1}^n y _k^{B'_{ik}} + \sum_{j=1}^n 
   L'_j \ln (y_j) 
\]
This demonstrates the result. Q.E.D. 

{\em Corollary 3.3:\/} The Poisson bracket of a GLVP system is form-invariant 
under QMTs. 

There is an important degree of freedom in the Poisson structure: 
Let $N \in \mbox{Ker} \{ K \}$. Then, the GLVP system we obtain does not 
change if we replace $L$ by $L+N$, since from (\ref{kdl}) we have 
$\lambda = K \cdot L = K \cdot (L+N)$. This is, in fact, a source of 
ambiguity in the Hamiltonian itself, because the flow is unaltered if we 
add to $H$ an extra term of the form:
\begin{equation}
\label{cnl}
   \phi _N = \sum_{j=1}^n N_j \ln (x_j) \; , \;\:\; N \in \mbox{Ker} \{ K \}
\end{equation}
This degree of freedom is precisely the one associated to the well-known 
Casimir functions: 

{\em Proposition 3.4:\/} Let $r=$ rank($K$) in a given GLVP structure. 
There is a complete set of $n-r$ functionally independent Casimir functions 
of the form (\ref{cnl}).

{\em Proof:\/} Evidently, rank(${\cal J}$) $=$ rank($K$) $=$ constant 
everywhere in the positive orthant. Then there are exactly $n-$rank($K$) 
functionally independent Casimirs. If $N \in \mbox{Ker} \{ K \}$ we have 
${\cal J} \cdot \nabla \phi _N = 0$, and all such $\phi _N$ are Casimirs. But 
dim(Ker$\{ K \}$) $= n-r$: Then we can get a maximal set of independent 
Casimirs by choosing $n-r$ linearly independent vectors of Ker$\{ K \}$, i.e. 
a basis of Ker$\{ K \}$. The functional independence can be readily verified 
in this case. Q.E.D.

Moreover, under a QMT of matrix $C$, $\phi _N$ is form-invariant and changes 
to $\phi _{N'}$, where $N'=C^T \cdot N$ and $N' \in 
\mbox{Ker} \{ K' \}$. Thus $\phi _{N'}$ is a Casimir of the 
transformed system. Let $\{ N^{(1)}, \ldots , N^{(n-r)} \}$ be a basis of 
Ker$\{K\}$, where $r$ $=$ rank($K$) as before. Then $\{ \phi_{N^{(1)}}, \ldots , 
\phi_{N^{(n-r)}}\}$ is a complete set of Casimirs of the initial system. 
After a QMT of matrix $C$, the Casimirs are transformed into a new family of 
Casimirs $\{ \phi_{N^{\prime \, (1)}}, \ldots , \phi_{N^{\prime \, (n-r)}}\}$, 
where $N^{\prime \, (i)}$ $=$ $C^T \cdot N^{(i)}$ for all $i$. In fact, the 
new set $\{ N^{\prime \, (1)}, \ldots , N^{\prime \, (n-r)} \}$ is also a 
basis of Ker$\{K'\}$. Therefore, every QMT carries a complete set of 
independent Casimirs of the form (\ref{cnl}) into its counterpart for the 
target system. In other words, the symplectic foliation of a GLVP system is a 
class property. 

We can equivalently express the Casimir functions (\ref{cnl}) in 
quasimonomial form as: 
\[
    \phi _N = \prod_{j=1}^n x_j^{N_j}  \; , \;\:\; N \in \mbox{Ker} \{ K \}
\]
However, we have seen in Proposition 2.4 that the quasimonomial first 
integrals arise in a purely GLV context, independently of the existence of a 
Poisson structure of the system. We saw that they are associated to a 
degeneracy in the rank of $M$. Clearly, there must be a close relationship 
between quasimonomial Casimirs and quasimonomial first integrals in general. 
We shall now demonstrate that both sets do coincide: 

{\bf Theorem  3.5:\/} Every quasimonomial constant of motion of a GLVP 
system is a Casimir function.

{\em Proof:\/} Notice that a quasimonomial constant of motion can be 
expressed as:
\[    
    \prod_{j=1}^n x_j^{N_j} = \mbox{constant} \; , \;\:\; N \in 
    \mbox{Ker} \{ M^T \}
\]
Therefore, our statement will be automatically demonstrated if we show that 
$\mbox{Ker} \{ M^T \}$ $=$ $\mbox{Ker} \{ K \}$. 

First, we demonstrate that rank($A$) $=$ rank($M$) for GLVP systems, where 
$M = (\lambda \mid A)$. From (\ref{kdl}), we can write symbolically $M = K 
\cdot ( L \mid B^T \cdot D)$. Expressed in this form, it is immediate to see 
that rank($A$) $=$ rank($M$) by simple inspection. Consequently, $\mbox{Ker} 
\{ M^T \}$ $=$ $\mbox{Ker} \{ A^T \}$ for GLVP systems, and the theorem will 
be demonstrated if $\mbox{Ker} \{ A^T \}$ $=$ $\mbox{Ker} \{ K \}$. 
 
It is evident that $\mbox{Ker} \{ K \} \subset \mbox{Ker} \{ A^T \}$. To show 
that $\mbox{Ker} \{ A^T \}$ $\subset$ $\mbox{Ker} \{ K \}$, note that 
rank($D \cdot B$) $=n$, and therefore Ker($D \cdot B$) $= \{ 0 \}$. The 
result is then straightforward, and the theorem is demonstated. Q.E.D.
                            
{\em Corollary 3.6:\/} In every GLVP system, rank($M$) $=$ rank($A$) $=$ 
rank(${\cal J}$) $=$ rank($K$). 

Theorem 3.5 will be of fundamental importance in what follows. It summarizes 
very well the interplay between algebraic and Poisson properties, which 
is present in GLVP systems. The level surfaces of the Casimir functions 
yield the global structure of the Poisson system ---they constitute the 
symplectic foliation of the phase-space. We 
are now able to reconstruct this important feature in terms of a very simple 
and purely algebraic scheme, which is summarized in the proof of Proposition 
2.4. More concisely, we can say that the symplectic foliation of the system 
is condensed in the rank properties of the GLV matrix $M$. Conversely, 
purely algebraic properties of the system in the GLV context now assume a 
completely new role. This is the case of the quasimonomial first integrals, 
which now become Casimir functions. Moreover, this parallelism is not only 
valid for a single system, but it is a class property. Obviously, some 
logical consequences arise from this interplay: For example, rank($M$) may 
take any value between 0 and $n$ in a general GLV system. However, from 
Corollary 3.6 this must be an even number for GLVP systems, an evident 
restriction if the level surfaces of the quasimonomial first integrals are 
to be a symplectic foliation. 

In Section II, we elaborated in some detail on the techniques for the 
manipulation of the quasimonomial constants of motion in a pure GLV 
context. These procedures involved inter-class transformations which might, 
if desired, eliminate all such first integrals. These ideas acquire 
completely new implications in the light of the last results: Such  
manipulations give now the key for restricting the dynamics to the symplectic 
leaves or, in the opposite sense, to embed the system in Poisson 
structures of higher dimensionality. We shall devote the next section to 
give a systematic treatment of these issues. 

\mbox{}

\begin{flushleft}
{\bf IV. TRANSFORMATIONS ON THE SYMPLECTIC FOLIATION} 
\end{flushleft}

According to Section II, three basic procedures are those which allow the 
manipulation of GLV systems: QMTs, embeddings, and decouplings. In Section 
III, we have seen how QMTs preserve the GLVP structure. That is, we 
have demonstrated that the GLVP structure is a class property. In this 
section, our aim will be to show that the inter-class operations do also 
preserve in an appropriate way the same GLVP character. This will complete 
the Poisson description of these systems and reinforce the parallelism 
between algebraic and Poisson properties. 

We shall first establish the result for the embeddings: 

{\em Proposition 4.1:\/} If a GLVP system which belongs to an 
$(r,n,m)$-class, with $m>n$, is subjected to a $p$-embedding, with  
$1 \leq p \leq m-n$, then the resulting system is also GLVP.

{\em Proof:\/} Relations $A=K \cdot B^T \cdot D$ and $\lambda = K \cdot L$ 
in the original system, imply that $\tilde{A}=\tilde{K} \cdot \tilde{B}^T 
\cdot \tilde{D}$ and $\tilde{\lambda} = \tilde{K} \cdot \tilde{L}$ in the 
embedded vector field, where:  
\[
    \tilde{K} = \left( \begin{array}{cc} 
                     K         &  O_{n \times p}   \\ 
               O_{p \times n}  &  O_{p \times p} 
                \end{array} \right) \; , \;\;\: 
    \tilde{L} = \left( \begin{array}{c}
                     L \\ L^*_{p \times 1}
                \end{array} \right) \; ,  \;\;\: 
    \tilde{D}=D \cdot E
\]
Here, $L^*$ is composed of arbitrary entries, $E$ is given by (\ref{e1}), 
and $\tilde{A}$, $\tilde{B}$, and $\tilde{ \lambda}$ are those in 
(\ref{bemb}). With the help of Theorem 3.1, this proves the 
result. Q.E.D. 

{\em Corollary 4.2:\/} Under the same hypotheses of Proposition 4.1, all the 
members of the $(r,n+p,m)$-class to which the expanded system belongs, are 
GLVP systems.

{\em Corollary 4.3:\/} Given a GLVP flow belonging to an $(r,n,m)$-class, 
every $m$-dimensional LV representative of the system is also GLVP. 

We shall demonstrate now that we have an analogous situation in the case of 
the decouplings. For this we need a preliminary result:

{\em Proposition 4.4:\/} Given a GLVP system of matrices $M^*$ and $K^*$, 
which belongs to an $(r,n+p,m)$-class, with $r \leq n$ and $0<p \leq m-n$, 
then there exists at least one QMT such that the transformed flow has 
matrices of the form:
\begin{equation}
\label{rpc}
   M = \left( \begin{array}{c}
            \bar{M} \\
            O_{p \times (m+1)}
        \end{array} \right)  \;,\:\:\;
   K = \left( \begin{array}{cc}
            \bar{K} & O_{n \times p} \\
            O_{p \times n} & O_{p \times p} 
        \end{array} \right) 
\end{equation}

{\em Proof:\/} Let $C$ be the matrix of the QMT. From the transformation rule 
$K = C^{-1} \cdot K^* \cdot (C^{-1})^T$ and the skew-symmetry of $K$, it is 
clear that there exists an invertible $C$ which recasts $K^*$ in the desired 
way. The form of $M$ is a consequence of the form of $K$, since from 
(\ref{kdl}) we have $\lambda$ $=$ $K \cdot L$ and $A$ $=$ 
$K \cdot B^T \cdot D$. Q.E.D. 

This leads to the main result for reductions:

{\em Proposition 4.5:\/} If a GLVP system belonging to an $(r,n+p,m)$-class, 
with $r \leq n$ and $0<p \leq m-n$, is subjected to a $p$-decoupling, then 
the resulting system is also GLVP. 

{\em Proof:\/} From Proposition 4.4, we can first transform the GLVP system 
into another one in the same $(r,n+p,m)$-class, with matrices like those in 
(\ref{rpc}). We shall denote the rest of matrices of the latter as $A$, $B$, 
$\lambda$, $L$ and $D$ ($A$ and $\lambda$ being given by equation 
(\ref{mcc})). We shall also, for convenience, express $B$ and $L$ 
as composed of submatrices in the usual form $B$ $=$ $(\bar{B} \mid 
\bar{B}'_{m \times p})$, and $L$ $=$ $( \bar{L} \mid \bar{L}'_{p 
\times 1})^T$. As we know from Section II, the matrices of the restricted 
system are $\hat{B}$ $=$ $\bar{B}$, $\hat{\lambda}$ $=$ $\bar{\lambda}$, 
and $\hat{A}$ $=$ $\bar{A} \cdot E^{-1}$, where $E$ is given by (\ref{e2}). 
It is not difficult to check that the relations $A=K \cdot B^T \cdot D$ and 
$\lambda = K \cdot L$, imply that $\hat{A}=\hat{K} \cdot \hat{B}^T \cdot 
\hat{D}$ and $\hat{\lambda} = \hat{K} \cdot \hat{L}$ in the reduced flow, 
where:  
\[
    \hat{K} = \bar{K} \; , \;\; \hat{L} = \bar{L} \; , \;\; \hat{D} = D 
    \cdot E^{-1} 
\]
This proves the result. Q.E.D.

Therefore, not only the GLV format itself, but also the proper GLVP structure 
is preserved when a reduction is carried out. This is consistent with the 
fact  that quasimonomial constants of motion are Casimirs of the Poisson 
structure. We can then state:

{\em Corollary 4.6:\/} If a GLV system of an $(r,n,m)$-class is GLVP, then 
all the $(r,k,m)$-classes, for $k = r, \ldots ,m$, which can be reached by 
means of successive embeddings and/or decouplings, are composed of GLVP 
systems. 

Consequently, all the transformations that we have defined within and among 
the classes preserve both the algebraic GLV properties of the equations and 
the Poisson format. A decoupling amounts to restricting the system totally or 
partially to the level surfaces of the Casimirs. An embedding adds new 
Casimirs to the system, thus increasing its dimensionality. The dynamics on 
the symplectic leaves remains, on the other hand, always intact. Knowing how 
to increase the dimension by means of embeddings is important, because it 
connects every class with the LV format ---and vice versa in the case of 
decouplings. In particular, this tells us how an LV system can be simplified. 

In the case in which we perform a maximal decoupling, we obviously 
arrive at an $(r,r,m)$-class of GLVP systems, with $r$ even: The members of 
this class are symplectic systems, since all the Casimirs have been removed. 
The previous results not only allow, however, the mere transformation of the 
vector field into symplectic form: They can also be used to reach the full 
reduction to the canonical forms of Darboux or Hamilton. This is the issue 
of the next section. 

\mbox{}

\begin{flushleft}
{\bf V. REDUCTION TO DARBOUX' CANONICAL FORM}
\end{flushleft}

We detail here three ways for constructing the Darboux' canonical form. We 
shall first address the more general approach, and then comment on two more 
specific possibilities. 

\mbox{}

\pagebreak
\noindent {\bf A. General method}

{\em Proposition 5.1:\/} Given a GLVP system belonging to an $(r,n,m)$-class, 
there exists a quasimonomial transformation such that for the transformed 
system:
\begin{equation}
\label{jdt}
    K' = {\cal S}(r,n-r) = \mbox{diag}(S_1,S_2, \ldots ,S_{r/2}, 
    \overbrace{ 0, \ldots ,0 }^{(n-r)}) \; , 
\end{equation}
where
\[  
  S_i = \left( \begin{array}{cc}
                      0  &  1 \\
                     -1  &  0
        \end{array} \right) \; , \;\;\; \forall \; i = 1, \ldots , r/2
\]

{\em Proof:\/} From Corollary 3.6, rank($K$) $=r$ for the original system. 
After a QMT of matrix $C$, we have from Proposition 3.2 that $K' = C^{-1} 
\cdot K \cdot (C^{-1})^T$, which is a congruence transformation over $K$. 
But, since $K$ is skew-symmetric and of rank $r$, it is congruent$^{38 \:}$ 
to a matrix of the form (\ref{jdt}). Then, the QMT exists. Q.E.D.

We can now state the following result: 

{\bf Theorem 5.2:} Every GLVP system belonging to an $(r,n,m)$-class can be 
globally reduced to Darboux' canonical form inside the positive orthant. 

{\em Proof:\/} The proof is constructive. We shall assume that the system has 
been already transformed in 
such a way that its matrix $K$ complies to format (\ref{jdt}). We can then 
introduce the following transformation, which is a global 
orientation-preserving diffeomorphism inside the positive orthant: 
\begin{equation}
\label{log}
    y_i = \ln (x_i) \; , \;\:\; i = 1, \ldots , n
\end{equation}
We now take into account the equation for the transformation of the structure 
matrix under general diffeomorphisms, $y_i = y_i(x)$:
\begin{equation}
\label{trnsj}
   ({\cal J'})_{ij}(y) = \sum_{k,l=1}^n \frac{\partial y_i}{\partial x_k}
   {\cal J}_{kl}(x) \frac{\partial y_j}{\partial x_l}
\end{equation}
The result, upon applying (\ref{log}), is:
\[
   ({\cal J'})_{ij} = ({\cal S}(r,n-r))_{ij} \; , \;\;\: 
   \forall \; i,j = 1, \ldots , n
\]
This transforms the original GLVP into a non-GLVP system that conforms, 
however, to Darboux' form: There are obviously $r/2$ pairs of canonically 
conjugate variables, and $(n-r)$ trivial Casimirs. The transformation of the 
Hamiltonian is straightforward. This proves the result. Q.E.D.

\mbox{}

\noindent {\bf B. Decoupling method}

Notice that in the case $r<n$, i.e. when the GLVP system is not symplectic, 
we can make use of the reduction procedure of the previous section, instead 
of applying Theorem 5.2 from the very beginning. The result would be another 
GLVP system, now symplectic, which belongs to an ($r,r,m$)-class. Making use 
of Theorem 5.2 on the reduced flow, would lead to a purely Hamiltonian 
system, since  
\[
   {\cal J'} = {\cal S}(r,0) \; , 
\]
which is the classical symplectic matrix of dimension $r$. 

\mbox{}

\noindent {\bf C. Linear transformation method}

Assume that we first subject the 
initial system of an $(r,n,m)$-class and matrix $K$, to transformation 
(\ref{log}). From (\ref{trnsj}), we find: 
\[
   {\cal J'} = K \; , 
\]
The resulting system $\dot{y}=K \cdot \nabla H'(y)$ is not GLVP, though it 
is Poisson. A linear change of variables $w=C \cdot y$, where $C$ is an 
invertible $n \times n$ matrix, leads to another Poisson system of constant 
structure matrix, 
\[
   \dot{w} = (C \cdot K\cdot C^T) \cdot \nabla H''(w)
\]
Finally, there exists a $C$ such that $C \cdot K\cdot C^T = {\cal S}(r,n-r)$, 
and Darboux' form is achieved. This procedure has already been applied in the 
literature to certain symplectic LV systems, of even-dimension and with a 
single fixed point.$^{37 \:}$ 

\mbox{}

\begin{flushleft}
{\bf VI. EXAMPLE: 3D LOTKA-VOLTERRA EQUATIONS}
\end{flushleft}

\noindent {\bf A. Poisson structure}

As an illustration of the previous results, we shall look upon the 3D LV 
Poisson structure first characterized by Nutku.$^{27 \:}$ The flow is 
given by the equations: 
\begin{eqnarray*}
   \dot{x}_1 &=& x_1( \rho + cx_2 +  x_3)   \\
   \dot{x}_2 &=& x_2( \mu  +  x_1 + ax_3)   \\
   \dot{x}_3 &=& x_3( \nu  + bx_1 +  x_2)   
\end{eqnarray*}
As Nutku has pointed out, this is a Poisson system if 
\[
   abc=-1 \; , \;\;\; \nu = \mu b - \rho ab
\]
In this case, the structure matrix and the Hamiltonian are, respectively: 
\[
       {\cal J} = \left( \begin{array}{ccc}
                   0     &   cx_1x_2   &   bcx_1x_3  \\
               -cx_1x_2  &     0       &   -x_2x_3   \\
               -bcx_1x_3 &   x_2x_3    &      0
                  \end{array} \right) \; \; ,
\]
obeying to form (\ref{J}), and
\[
   H = abx_1+x_2-ax_3+ \nu \ln (x_2) - \mu \ln (x_3)
\]
which is of the form (\ref{H}). The system is thus GLVP with characteristic 
matrices:
\[
   B = I_{3 \times 3} \; , \;\;\; 
   M = \left( \begin{array}{cccc}
         \rho & 0 & c & 1 \\
         \mu  & 1 & 0 & a \\
         \nu  & b & 1 & 0
       \end{array} \right) \; , 
\]
\[
   K = \left( \begin{array}{ccc}
          0 &  c & bc \\
         -c &  0 & -1 \\
        -bc &  1 & 0 
       \end{array} \right)
\; , \;\;\; 
   D = \left( \begin{array}{ccc}
         ab &  0 &  0 \\
         0  &  1 &  0 \\
         0  &  0 & -a
       \end{array} \right)
\; , \;\;\; 
   L = \left( \begin{array}{c}
         0 \\ \nu \\ - \mu 
       \end{array} \right)
\]
Notice that rank($M$) $=$ rank($A$) $=$ rank($K$) $=$ rank(${\cal J}$) $=$ 2 
inside the positive orthant of R$^3$. There is then one independent Casimir 
function. By noting that, in $M$, row(3) $=$ $(1/c) \times$row(1) $+$ 
$b \times$row(2), we immediately find the quasimonomial first integral: 
\[
    x_1^{ab}x_2^{-b}x_3 = \mbox{constant} \; \; , 
\]
which is also a Casimir of the Poisson structure from Theorem 3.5. This way 
of recovering Casimir functions of the system is certainly more economic 
than solving the PDE $\; {\cal J} \cdot \nabla \phi = 0$, which is the usual 
approach. 

\mbox{}

\noindent {\bf B. Darboux' form: General method}

Let us now subject the system to a QMT of matrix: 
\[
   C = \left( \begin{array}{ccc}
          c & 0 & 0  \\
          0 & 1 & 0  \\
          1 & b & -1 
       \end{array} \right)
\]
We arrive to a new GLVP of matrices:
\[
   B' = \left( \begin{array}{ccc}
          c & 0 & 0  \\
          0 & 1 & 0  \\
          1 & b & -1 
       \end{array} \right) \; , \;\;\; 
   M' = \left( \begin{array}{cccc}
         \rho /c & 0 & 1 & 1/c \\
         \mu  & 1 & 0 & a \\
          0  & 0 & 0 & 0
       \end{array} \right) \; , 
\]
\[
   K' = \left( \begin{array}{ccc}
          0 & 1 & 0 \\
         -1 & 0 & 0 \\
          0 & 0 & 0 
       \end{array} \right)
\; , \;\;\; 
   D' = \left( \begin{array}{ccc}
         ab &  0 &  0 \\
         0  &  1 &  0 \\
         0  &  0 & -a
       \end{array} \right)
\; , \;\;\; 
   L' = \left( \begin{array}{c}
         - \mu \\ \rho /c \\ \mu 
       \end{array} \right)
\]
The Casimir function has been decoupled, and now is just $x'_3$. A change of 
variables $y_i$ $=$ $ \ln (x'_i)$, $i$ $=$ $1,2,3$ yields Darboux's form, 
with:
\begin{equation}
\label{jdf3d}
    {\cal J}(y) = \left( \begin{array}{ccc}
                   0 & 1 & 0 \\ 
                  -1 & 0 & 0 \\ 
                   0 & 0 & 0 
                  \end{array} \right) 
\end{equation}
and
\begin{equation}
\label{hdf3d}
   H(y) = ab e^{cy_1}+e^{y_2}-ae^{y_1+by_2-y_3} - \mu y_1 + ( \rho /c)y_2 + 
   \mu y_3 
\end{equation}

\mbox{}

\noindent {\bf C. Darboux' form: Decoupling method}

Although the previous one is the shortest way to achieve the transformation 
into Darboux' form, it may be sometimes more convenient to proceed in a 
two-step alternative: The first step is the transformation of the system 
into a symplectic flow. This might be more appropriate in systems of higher 
dimensions, in which an initial reduction of the dimensionality of the 
problem may produce the most manageable system. We shall briefly display it 
for the sake of illustration. 

We can first make a QMT of matrix:
\[
   C_1 = \left( \begin{array}{ccc}
          1   & 0 & 0  \\
          0   & 1 & 0  \\
          1/c & b & -1 
       \end{array} \right)
\]
If we then decouple the third variable, assuming for simplicity that its 
initial condition is $x'_3(0)=1$, the reduced GLVP system is given by: 
\[
   \hat{B}' = \left( \begin{array}{cc}
               1 & 0 \\ 0 & 1 \\ 1/c & b 
             \end{array} \right) \; , \;\;\; 
   \hat{M}' = \left( \begin{array}{cccc}
         \rho & 0 & c & 1 \\
         \mu  & 1 & 0 & a 
       \end{array} \right) \; , 
\]
\[
   \hat{K}' = \left( \begin{array}{cc}
          0 &  c \\
         -c &  0 
       \end{array} \right)
\; , \;\;\; 
   \hat{D}' = \left( \begin{array}{ccc}
         ab &  0 &  0 \\
         0  &  1 &  0 \\
         0  &  0 & -a
       \end{array} \right)
\; , \;\;\; 
   \hat{L}' = \left( \begin{array}{c}
         - \mu /c \\ \nu - \mu b 
       \end{array} \right)
\]
We now perform a second QMT, this time acting on the reduced $(2,2,3)$-class:
\[
   C_2 = \left( \begin{array}{cc}
          c & 0  \\
          0 & 1   
       \end{array} \right)
\]
In the resulting flow, we have:
\[
   \hat{K}'' = S(2,0) = 
      \left( \begin{array}{cc}
           0 & 1  \\
          -1 & 0   
       \end{array} \right)
\]
After the final change of variables $y_i = \ln (x''_i)$, $i=1,2$, we arrive 
at a Hamiltonian system in which:
\[
   H(y) = ab e^{cy_1}+e^{y_2}-ae^{y_1+by_2} - \mu y_1 + ( \rho /c)y_2 
\]
Notice that this Hamiltonian can be obtained from (\ref{hdf3d}) with $y_3=0$. 
This is due to the initial condition we have assumed for $x'_3$ in the 
1-decoupling. We obviously retrieve, up to trivial differences in form, the 
Darboux system (\ref{jdf3d}-\ref{hdf3d}). 

\mbox{}

\begin{flushleft}
{\bf VII. FINAL REMARKS}
\end{flushleft}

We have seen that there is a close parallelism between the Poisson 
structure of GLVP flows and the algebraic properties of GLV equations. 
The deep and unexpected interplay between both aspects of the systems 
results in an economy in their description. It also establishes an 
operational framework for their manipulation and simplification. This is, to 
our knowledge, a novel approach to the treatment of finite-dimensional 
Poisson structures. 

We end this work by giving an evaluation in relation to what has been done 
previously in the literature. Unfortunately, the only way for doing this 
is by particularizing the comparison to LV models, for which earlier results 
are available. We shall only consider previous approaches which are valid for 
$n$-dimensional LV models, for arbitrary $n$. We can then say that most 
Hamiltonian and Poisson LV systems treated in the literature have a matrix 
$A$ which is of maximal rank, i.e., they have a single fixed 
point.$^{6,30,37 \:}$ Also, they are often restricted to 
even dimensionality,$^{30,37 \:}$ which usually entails that the 
system is symplectic (these are, of course, two requirements which are 
implicit in the classical Hamiltonian studies). None of these restrictions 
is present, as we have seen, in our models. On the other hand, our treatment 
joins previous works in what concerns certain requirements on matrix $A$: 
For a GLVP Lotka-Volterra system, we have that $A=K \cdot D$, where $K$ is 
skew-symmetric. This implies, as it can be readily seen, that $D_{ii} A_{ij}$ 
$=$ $-D_{jj} A_{ji}$, which is exactly the same kind of generalized 
skew-symmetry which can be found in the works of Kerner$^{37 \:}$ and 
some cases from Plank,$^{6 \:}$ for example. Therefore, the scope of our 
treatment does not differ, in this sense, to that of previous ones. Notice 
also how our Hamiltonian (\ref{H}) reduces, in the case of LV systems, to a 
generalization of the classical Volterra's constant of the motion:
\[
    H_V = \sum_{i=1}^n \beta _i (x_i - p_i \ln (x_i)) \;\;,
\]
where $p_i$ are the coordinates of the (unique) fixed point of Volterra's 
systems. 

Another interesting issue which we would like to comment here concerns the 
use of an arbitrary Hamiltonian, while retaining the form (\ref{J}) for the 
structure matrix. This leads, of course, to the generation of a wide range 
of Poisson systems. This procedure can be found, for example, in Plank's 
work.$^{6 \:}$ We may mention that many of our previous results still hold in 
this Hamiltonian-independent situation. This is the case, for instance, in 
the reduction of the system to the Darboux' form. The reason is that the 
criterion to decide whether a system complies to Darboux' format or not, 
relies on the form of the structure matrix, exclusively. Consequently, the 
manipulations to which the system is to be subjected concern the recasting 
of ${\cal J}$ in the desired form, $H$ being irrelevant for that case 
---which is the situation in Section V. 

\mbox{}

\begin{flushleft}
{\bf Acknowledgements}
\end{flushleft}

This work has been supported by the DGICYT of Spain (grant PB94-0390) and by 
the E.U. (Esprit WG 24490). B. H. acknowledges a doctoral fellowship from 
Comunidad de Madrid. 

\mbox{}

\begin{flushleft}
{\bf References and notes}
\end{flushleft}
   $^1$ A. Lichnerowicz, J. Diff. Geom. {\bf 12}, 253 (1977); A. Weinstein, 
      J. Diff. Geom. {\bf 18}, 523 (1983). \newline 
   $^2$ P. J. Olver, {\em Applications of Lie Groups to 
      Differential Equations\/} (Springer-Verlag, New York, 1993), 2nd ed. 
      \newline 
   $^3$ H. G\"{u}mral and Y. Nutku, J. Math. Phys. {\bf 34}, 5691 
      (1993); F. Haas and J. Goedert, Phys. Lett. A {\bf 199}, 173 (1995). 
      \newline 
   $^4$ S. A. Hojman, J. Phys. A: Math. Gen. {\bf 24}, L249 (1991); 
      S. A. Hojman, J. Phys. A: Math. Gen. {\bf 29}, 667 (1996). \newline 
   $^5$ M. Plank, Nonlinearity {\bf 9}, 887 (1996). \newline 
   $^6$ M. Plank, J. Math. Phys. {\bf 36}, 3520 (1995). \newline 
   $^7$ J. Goedert, F. Haas, D. Hua, M. R. Feix and L. Cair\'{o}, 
      J. Phys. A: Math. Gen. {\bf 27}, 6495 (1994). \newline 
   $^8$ W. Pauli, Nuovo Cimento {\bf 10}, 648 (1953). \newline 
   $^9$ J. Gibbons, D. D. Holm and B. Kupershmidt, Phys. Lett. A 
      {\bf 90}, 281 (1982); D. D. Holm and B. A. Kupershmidt, Phys. Lett. A 
      {\bf 91}, 425 (1982); D. D. Holm and B. A. Kupershmidt, Phys. Lett. A 
      {\bf 93}, 177 (1983); J. E. Marsden, R. Montgomery, P. J. Morrison and 
      W. B. Thompson, Ann. Physics {\bf 169}, 29 (1986); D. Chinea, J. C. 
      Marrero and M. de Le\'{o}n, J. Phys. A: Math. Gen. {\bf 29}, 6313 
      (1996). \newline 
   $^{10}$ D. D. Holm, Physica D {\bf 17}, 1 (1985); D. Lewis, J. 
      Marsden, R. Montgomery and T. Ratiu, Physica D {\bf 18}, 391 (1986). 
      \newline 
   $^{11}$ P. J. Morrison and J. M. Greene, Phys. Rev. Lett. {\bf 45}, 
      790 (1980). \newline 
   $^{12}$ D. D. Holm, Phys. Lett. A {\bf 114}, 137 (1986). \newline 
   $^{13}$ J. E. Marsden and A. Weinstein, Physica D {\bf 4}, 394 
      (1982). \newline 
   $^{14}$ R. D. Hazeltine, D. D. Holm and P. J. Morrison, J. Plasma 
      Physics {\bf 34}, 103 (1985). \newline 
   $^{15}$ D. D. Holm and B. A. Kupershmidt, Physica D {\bf 6}, 347 
      (1983). \newline 
   $^{16}$ I. E. Dzyaloshinskii and G. E. Volovick, Ann. Physics 
      {\bf 125}, 67 (1980). \newline 
   $^{17}$ R. G. Littlejohn, J. Math. Phys. {\bf 20}, 2445 (1979); 
       R. G. Littlejohn, J. Math. Phys. {\bf 23}, 742 (1982); J. R. Cary and 
       R. G. Littlejohn, Ann. Physics {\bf 151}, 1 (1983). \newline 
   $^{18}$ H. D. I. Abarbanel, D. D. Holm, J. E. Marsden and T. Ratiu, 
      Phys. Rev. Lett. {\bf 52}, 2352 (1984); D. D. Holm, J. E. Marsden, T. 
      Ratiu and A. Weinstein, Phys. Rep. {\bf 123}, 1 (1985); D. Lewis, J. 
      Marsden and T. Ratiu, J. Math. Phys. {\bf 28}, 2508 (1987); D. D. Holm 
      and K. B. Wolf, Physica D {\bf 51}, 189 (1991).  \newline 
   $^{19}$ J. C. Simo, T. A. Posbergh and J. E. Marsden, Phys. Rep. 
      {\bf 193}, 279 (1990); J. C. Simo, D. Lewis and J. E. Marsden, Arch. 
      Rational Mec. Anal. {\bf 115}, 15 (1991); J. C. Simo, T. A. Posbergh 
      and J. E. Marsden, Arch. Rational Mec. Anal. {\bf 115}, 61 (1991). 
      \newline 
   $^{20}$ D. David, D. D. Holm and M. V. Tratnik, Phys. Lett. A 
      {\bf 137}, 355 (1989); D. David, D. D. Holm and M. V. Tratnik, Phys. 
      Lett. A {\bf 138}, 29 (1989); D. David, D. D. Holm and M. V. Tratnik, 
      Phys. Rep. {\bf 187}, 281 (1990). \newline 
   $^{21}$ F. Magri, J. Math. Phys. {\bf 19}, 1156 (1978); P. J. Olver, 
      Phys. Lett. A {\bf 148}, 177 (1990).  \newline 
   $^{22}$ D. David and D. D. Holm, J. Nonlinear Sci. {\bf 2}, 241 
      (1992). \newline 
   $^{23}$ P. J. Morrison and R. D. Hazeltine, Phys. Fluids {\bf 27},  
      886 (1984).  \newline 
   $^{24}$ R. G. Littlejohn, AIP Conf. Proc. {\bf 88}, 47 (1982). \newline 
   $^{25}$ L. Cair\'{o} and M. R. Feix, J. Phys. A: Math. Gen. {\bf 25}, 
      L1287 (1992). \newline 
   $^{26}$ M. Razavy and F. J. Kennedy, Can. J. Phys. {\bf 52}, 1532 
      (1974); C. A. Lucey and E. T. Newman, J. Math. Phys. {\bf 29}, 2430 
      (1988); Y. Nutku, J. Phys. A: Math. Gen. {\bf 23}, L1145 (1990).  
      \newline 
   $^{27}$ Y. Nutku, Phys. Lett. A {\bf 145}, 27 (1990). \newline 
   $^{28}$ B. Hern\'{a}ndez--Bermejo and V. Fair\'{e}n, Phys. Lett. A 
      {\bf 234}, 35 (1997). \newline 
   $^{29}$ A. J. Lotka, {\em Elements of Mathematical Biology\/} (Dover, 
      New York, 1956). \newline 
   $^{30}$ V. Volterra, {\em Le\c{c}ons sur la Th\'{e}orie 
      Math\'{e}matique de la Lutte pour la Vie\/} (Gauthier Villars, Paris, 
      1931). \newline 
   $^{31}$ E. H. Kerner, Bull. Math. Biophys. {\bf 26}, 151 (1964); E. H. 
      Kerner, in {\em Advances in Chemical Physics,\/} edited by I. Prigogine 
      and S. A. Rice (Wiley, New York, 1971), Vol. 19; N. S. Goel, S. C. 
      Maitra and E. W. Montroll, Rev. Mod. Phys. {\bf 43}, 231 (1971); E. H. 
      Kerner, {\em Gibbs Ensemble: Biological Ensemble\/} (Gordon and Breach, 
      New York, 1972).  \newline 
   $^{32}$ L. Cair\'{o} and M. R. Feix, J. Math. Phys. {\bf 33}, 2440 
      (1992). \newline 
   $^{33}$ M. Peschel and W. Mende, {\em The Predator--Prey Model. Do 
      we live in a Volterra World?\/} (Springer-Verlag, Vienna--New York, 
      1986).  \newline 
   $^{34}$ L. Brenig, Phys. Lett. A {\bf 133}, 378 (1988); L. Brenig and 
      A. Goriely, Phys. Rev. A {\bf 40}, 4119 (1989); A. Br'uno, {\em Local 
      Methods in the Theory of Differential Equations\/} (Springer-Verlag, 
      New York, 1989); J. L. Gouz\'e, {\em Transformation of polynomial 
      differential systems in the positive orthant\/} (Rapport INRIA No. 
      1308, Sophia-Antipolis, 06561 Valbonne Cedex, France, 1990). \newline 
   $^{35}$ A. Goriely and L. Brenig, Phys. Lett. A {\bf 145}, 245 
      (1990); A. Goriely, J. Math. Phys. {\bf 33}, 2728 (1992); B. 
      Hern\'{a}ndez--Bermejo and V. Fair\'{e}n, Phys. Lett. A {\bf 206}, 31 
      (1995); V. Fair\'{e}n and B. Hern\'{a}ndez--Bermejo, J. Phys. Chem. 
      {\bf 100}, 19023 (1996); B. Hern\'{a}ndez--Bermejo and V. Fair\'{e}n, 
      Math. Biosci. {\bf 140}, 1 (1997). \newline 
   $^{36}$ L. Brenig and A. Goriely, in {\em Computer Algebra and 
      Differential Equations,\/} edited by E. Tournier (Cambridge U. P., 
      Cambridge, England, 1994). \newline 
   $^{37}$ E. H. Kerner, Phys. Lett. A {\bf 151}, 401 (1990); E. H. 
      Kerner, J. Math. Phys. {\bf 38}, 1218 (1997).  \newline 
   $^{38}$ F. Ayres Jr., {\em Schaum's Outline of Matrices\/} 
      (McGraw-Hill, New York, 1962).  
\end{document}